# Efficient Data Gathering in Wireless Sensor Networks Based on Matrix Completion and Compressive Sensing

Jiping Xiong, Member, IEEE, Jian Zhao and Lei Chen

***Abstract*—Gathering data in an energy efficient manner in wireless sensor networks is an important design challenge. In wireless sensor networks, the readings of sensors always exhibit intra-temporal and inter-spatial correlations. Therefore, in this letter, we use low rank matrix completion theory to explore the inter-spatial correlation and use compressive sensing theory to take advantage of intra-temporal correlation. Our method, dubbed MCCS, can significantly reduce the amount of data that each sensor must send through network and to the sink, thus prolong the lifetime of the whole networks. Experiments using real datasets demonstrate the feasibility and efficacy of our MCCS method.**

## I. Introduction

IN wireless senor networks, deployed sensor nodes periodically collect readings and send them to sinks (or base stations) via wireless channels. Due to the limited computation capability and energy power, it is desirable to design simple and energy efficient data gathering method to reduce energy consumption on each sensor.

Energy consumption of sending and receiving data is one of the most important factors in WSNs. Various methods have been proposed to achieve data compression include tradition source coding and distributed source coding [1]. Beside those methods, another promising taxonomy is Compressive Sensing (CS)[2-3] based approaches.

CS theory proves that if discrete signal $x_N \in R^N$ can be sparsely represented as $q_N$ using a transform basis $\Psi_{N\times N}$ (e.g. wavelet basis), we can exactly recover $x_N$ from $y_M$ (M<N) measurements using $l1$ minimization constrain, here

$$y_M = \Phi_{M\times N} x_N = \Phi_{M\times N} \Psi_{N\times N} q_N \qquad (1)$$

In (1), $\Phi_{M\times N}$ is called the sensing matrix. In this letter, we will use sparse binary matrix as the sensing matrix which can reduce energy consumption while achieving competitive data compression ratio [4]. $q_N$ is a sparse vector with only few non-zero elements.

Vuran et al.[5] pointed out that in wireless sensor networks, the phenomena observed by sensors is highly spatially and temporally correlated. Those correlations result to the sparsity of sensor readings under wavelet transform (see fig. 1(b) for one example of real world dataset) which then satisfy the sparse requirement of CS theory.

In [6-7] both inter-signal and intra-signal correlations are considered to reduce the communication cost on each sensor. Each sensor sends CS compressed measurements to sink individually, and sink recovers origin readings from measurements of all the sensors using Joint Spare Module (JSM).

In [8] the authors proposed Compressive Data Gathering (CDG) framework to make use of the sparsity of inter-signal in multi-hop fashion for large dense wireless sensor networks. Also in [9], a Hybrid CS aggregation was proposed to further reduce the communication overhead of sensors.

Temporal correlation of intra-signal and hierarchical cluster feature were exploited in [10], cluster heads of each layer perform CS reconstruction and use the recovered readings to form new CS measurements with shorter length.

Matrix Completion (MC) is the theory to recover of a full data matrix from part of its entries. Recently, candès et al. proves that if the data matrix is a low rank or approximately low rank matrix, we can recover missing entries from an incomplete set of entries [11]. Observing that the readings of sensors can form approximately low rank data matrix [5], Cheng et al. proposed EDCA scheme to apply the low rank matrix completion theory to the data gathering problem of WSNs [12]. Since EDCA only sample partial readings on each single sensor, the energy cost of sampling which is often ignored before can also be reduced.

Inspired by the low rank matrix completion theory and intra-signal temporal correlation, in this letter we first combine the matrix completion and compressive sensing theory to further compress the senor readings.

Manuscript received XX X, 2013. This work was supported in part by the Opening Fund of Top Key Discipline of Computer Software and Theory in Zhejiang Provincial Colleges at Zhejiang Normal University.

Jiping Xiong and Jian Zhao are with the College of Mathematics,Physics and Information Engineering, Zhejiang Normal University (corresponding e-mail: jpxiong@ieee.org).

Lei Chen is with Broadband Wireless Communications and Multimedia Laboratory, School of Electronics and Information, TongJi University (e-mail: lchen.tongji@gmail.com).

This letter first presents our matrix completion and compressive sensing method, which are called MCCS, and then demonstrates the performance gains by real deployed dataset. At last, we conclude and give out future research direction in Section IV.

## II. Proposed Method

### A. Encoding Algorithm at Each Sensor Node

Our aim is to try to reduce the energy cost in all possible aspects including sample energy cost, communication cost and computation complex. Below are the details of our encoding algorithm at each sensor.

*Step 1*: Each sensor node randomly generates a binary sampling position vector $P_N$ with only q (q<N) non-zero entries. Here, we call $l = \lfloor N/q \rfloor$ the MC (Matrix Completion) sample ratio.

*Step 2*: Each sensor node then scans the binary sampling position vector and only samples when the corresponding entry is non-zero. At the end, those sampled readings form a vector $x_q \in R^q$.

At this step, MC based compression is applied. Each node only sample q readings instead of N which results to $l = \lfloor N/q \rfloor$ compression ratio. Thus the energy cost of sampling is reduced.

*Step 3*: At the first time, each sensor node generates and stores the same sparse binary matrix $B_{p\times q}$ (p<q) using the seed K pre-shared among all sensors and sinks. There are only small number d ( $p > d \geq 1$) non-zero elements random located in each column of $B_{p\times q}$ . Also, we call $h = [p/q]$ the CS compression ratio.

*Step 4:* Each sensor node gets CS measurements $y_p$ from $x_q$ according below operation

$$y_p = B_{p\times q} \times x_q \tag{2}$$

After sending out $y_p$ and the sampling position vector $P_N$ (using bit mode) will be sent out, goes back to step 1.

At this step, since our measurement matrix is a sparse binary matrix, the energy expenditure for this CS compression is only involves simple addition operation.

Although $x_q$ is randomly sampled from N continuous readings, due to the temporal correlation, we find that $x_q$ can still be sparse under certain transform basis (fig. 1(c) is an example from a real dataset). That is why we can still using CS to compress the readings after Matrix Completion based compression.

At the end, each sensor only needs to send out p readings instead of N, which results to $p/N = l \times h$ total compression ratio.

### B. Recovering Algorithm at Sink Node

At the beginning, sink node generates and stores the same sparse binary matrix $B_{p\times q}$ using shared seed K.

After receiving CS measurements $y_p$ from each node, sink node is able to reconstruct the partial readings $x_q$ through solving a $l_1$ minimization problem:

$$\min \|q_q\|_{l_1} \quad s.t. \quad y_p = B_{p\times q} x_q, x_q = \Psi_{q\times q} q_q \tag{3}$$

Here $\Psi_{N\times N}$ is a transform basis that can make $x_q$ sparsely represented as $q_q$ . Suppose $\hat{q}_q$ is the solution to the convex optimization problem, than the original partial readings is $\hat{x}_q = \Psi_{q\times q}\hat{q}_q$ . With proper values of p and q, the error between $\hat{x}_q$ and $x_q$ can be very small.

After recovering from CS compression, sink node uses $\hat{x}_q$ and the binary sampling position vector $P_N$ from each sensor node forming an incomplete readings matrix $\hat{X}_{J\times N}$ where J is the number of sensor nodes. According to the spatial correlation, $\hat{X}_{J\times N}$ is an approximate low rank matrix. That means we can recover the full readings matrix from convex optimization problem [11]:

$$\begin{aligned} &\text{minimize } \|X\|_* \\ &\text{subject to } X_{ij} = \hat{X}_{ij} \quad (i,j) \in \Omega \end{aligned} \tag{4}$$

Here, $\|.\|_*$ is the nuclear norm, and the set $\Omega$ is the locations corresponding to the partially sampled sets of readings.

## III. Experiment

We use the real world dataset from Intel Berkeley Research Lab [11] to evaluate the efficiency of our proposed method. The dataset contains temperature, humidity, light and voltage value periodically collected every 31 seconds from 54 distributed sensor nodes between February 28th and April 5th, 2004. In our experiment, temperature values from 46 nodes on March 1st, 2004 are selected (eight nodes among 54 nodes have very few values). Thus, those traces form a matrix $X_{46\times 250}$ , here each row is from the readings of each single sensor.

$X_{46\times 250}$ is not row rank but approximately low rank. Indeed, letting $X^1_{46\times 250}$ be the rank-1 approximation, we have $\|X^1_{46\times 250} - X_{46\times 250}\|_F / \|X_{46\times 250}\|_F = 0.0068$, here $\|.\|_F$ is the Frobenius norm.

Fig. 1(a) is the original readings from 10th row of $X_{46\times 250}$ which belongs to sensor node 10. Fig. 1(b) shows the 250 coefficients after 6 level 9/7 wavelet de-correlation. There are only 5 coefficients whose absolute value is larger than 0.05, which means those readings can be sparse in wavelet domain. More importantly, fig. 1(c) shows that even after randomly sampling from original readings ( $l$ =0.5), those partial sampled readings can also be sparse in wavelet domain.

In this experiment, we use CVX toolbox [14] doing CS reconstruction and TFOCS toolbox [15] doing matrix completion recovering.

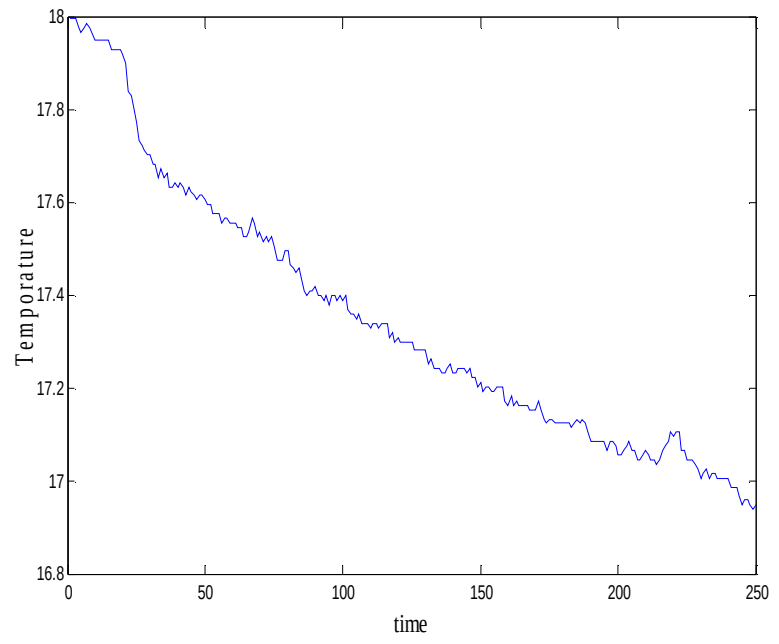

(a) Origin data of sensor node 10

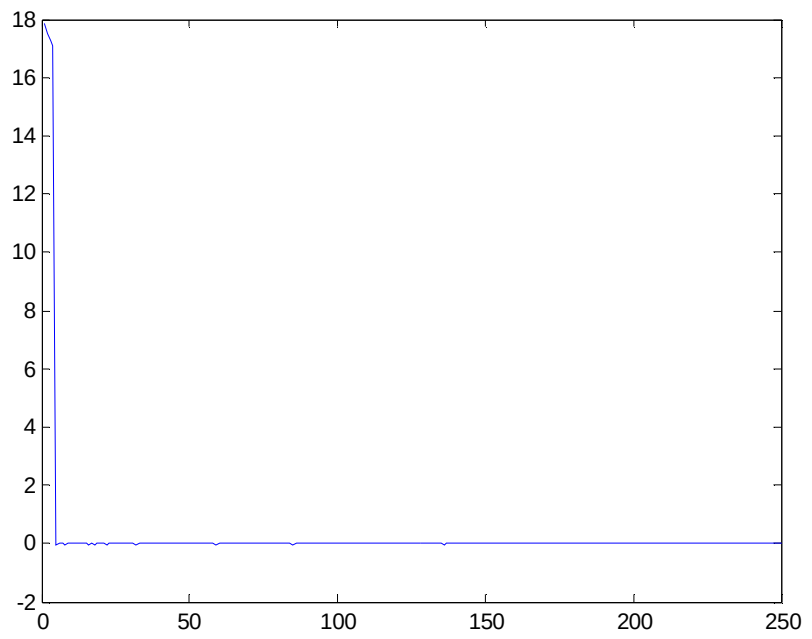

(b) Coefficients after 6-level 9/7 wavelet transform

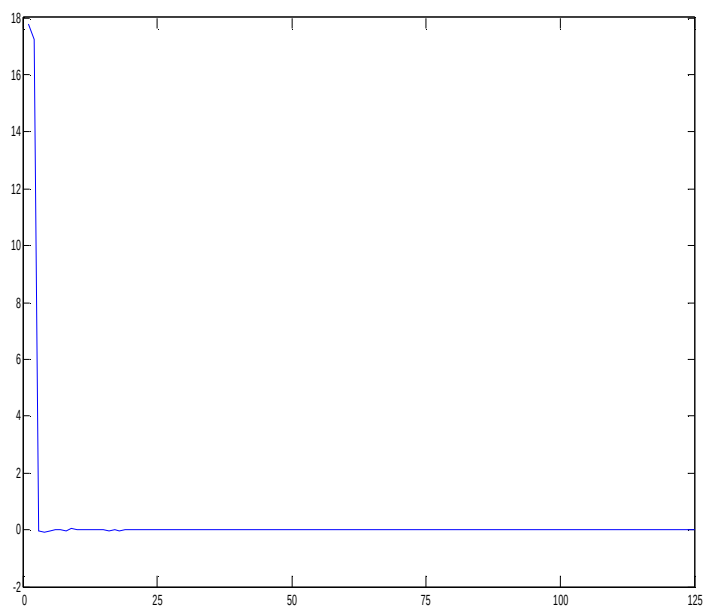

(c) Coefficients after 6-level 9/7 wavelet transform of randomly selected data with $l$ =0.5.

Fig. 1. DCT transform of origin sensor readings for node 10.

The number of iteration of simulation is 100, and we calculate the average result values. In each simulation, we calculate an error matrix by comparing the recovered matrix with the original matrix $X_{46\times250}$. Fig. 2 shows the mean value and standard deviation on different sample ratio $l$ between MC method and our MCCS method. From the results, we can see that with almost the same accurate, we can achieve more compression ratio than MC method. For example, at the same 0.3 sample ratio, if we set CS compression ratio $h$ to 0.4, our total compression ratio is 0.12 which is much less than MC method.

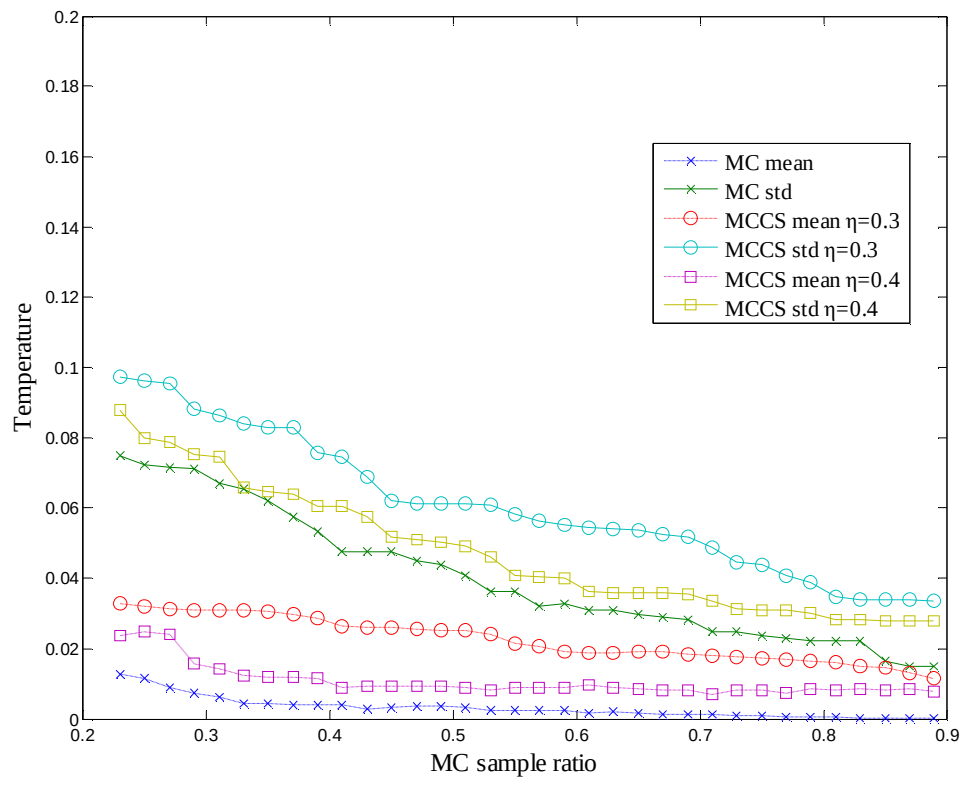

Fig. 2. Deviation and mean value with different sample ratio

## IV. Conclusion

In this letter, we propose a Matrix Completion and Compressive Sensing based data gathering method in wireless sensor networks in a computationally and energy efficient manner for sensor nodes. Our method fully exploits the low rank and sparse nature of readings among sensor nodes, and can achieve a high compression ratio of readings. Experiment on real world dataset demonstrates the efficient of the method.

Our future work include: 1) do more experiments on other WSN deployment datasets using other CS reconstruction algorithms and MC recovering algorithms; 2) apply DCS-OMP [7] to the CS reconstruction procedure to see whether we can further reduce the number of samples required for effective readings recovery; and 3) modify our method to accommodate to abnormal readings.

## References

[1] Marco F. Duarte, Godwin Shen, Antonio Ortega and Richard G. Baraniuk, “Signal compression in wireless sensor networks,” *Phil. Trans. R.Soc.A*, Jan 2012, pp. 118-135.

[2] D.L.Donoho, “Compressed Sensing,” *IEEE Transactions. Information Theory*, vol.52, no.4, April 2006, pp. 1289-1306.

[3] E.J.Candes, “Compressive Sampling,” *International Congress of Mathematicians*, vol.3, pp. 1433-1452, Madrid, Spain, 2006.

[4] R.Berinde, P.Indyk, “Sparse recovery using sparse random matrices,” MIT-CSAIL Technical Report, 2008

[5] MC Vuran, ÖB Akan, IF Akyildiz, “Spatio-temporal correlation: theory and applications for wireless sensor networks,” *Computer Networks*, Vol.45, Issue.3, 2004, pp.245-259.

[6] M. F. Duarte, S. Sarvotham, M. B. Wakin, D. Baron, and R. G. Baraniuk, “Joint Sparsity Models for Distributed Compressed Sensing,” *In Proceedings of the Workshop on Signal Processing with Adaptive Sparse Structured Representations*, 2005.

[7] D.Baron, F.M.Duarte, M.B.Wakin, S.Sarvotham, and R.G.Baraniuk, “Distributed compressive sensing,” *Sensor, Signal and Information Processing (SenSIP) Workshop*, 2008

[8] C. Luo, F. Wu, J. Sun, and C. W. Chen, “Compressive data gathering for large-scale wireless sensor networks,” In Proceedings of MobiCom, 2009.

[9] Liu Xiang, Jun Luo, Athanasios Vasilakos, “Compressed Data Aggregation for Energy Efficient Wireless Sensor Networks”, *8th Annual IEEE Communications Society Conference on Sensor, Mesh and Ad Hoc Communications and Networks*, 2011, pp46-54.

[10] Xi Xu, Rashid Ansari, Ashfaq Khokhar, “Power-efficient Hierarchical Data Aggregation using Compressive Sensing in WSN,” Oct 2012, *arXiv*: 1210.3876v1, pp. 1-6.

[11] E. Candès and B. Recht, “Exact Matrix Completion via Convex Optimization,” *Foundations of Computational Mathematics*, Vol.9, 2009, pp. 717-772.

[12] J. Cheng, H. Jiang, X. Ma, L. Liu, L. Qian, C. Tian, and W. Liu. “Efficient Data Collection with Sampling in WSNs: Making Use of Matrix Completion Techniques,” *In Proc. of IEEE GLOBECOM*, Dec 2010, pp:1-5.

[13] Intel Berkeley Research Lab Data, [Online]. Available: http://db.csail.mit.edu/labdata/labdata.html

[14] CVX Research, Inc. CVX: Matlab software for disciplined convex programming, version 2.0 beta. http://cvxr.com/cvx, September 2012.

[15] TFOCS: Templates for First-Order Conic Solvers, http://cvxr.com/tfocs/ , August 2011.